\documentclass[conference]{IEEEtran}
\IEEEoverridecommandlockouts
\usepackage{cite}
\usepackage{amsmath,amssymb,amsfonts}
\usepackage{algorithmic}
\usepackage{graphicx}
\usepackage{textcomp}
\usepackage{xcolor}
\usepackage{verbatim}
\usepackage{hyperref}
\usepackage{url}
\usepackage{amsmath}
\def\BibTeX{{\rm B\kern-.05em{\sc i\kern-.025em b}\kern-.08em
    T\kern-.1667em\lower.7ex\hbox{E}\kern-.125emX}}

\begin{document}

\title{Component Based Quantum Machine Learning Explainability}
\author{
  \begin{minipage}{\textwidth}
    \centering
    {\Large Barra White $^1$, Krishnendu Guha $^2$}\\
    \textit{School of Computer Science and Information Technology}\\
    \textit{University College Cork}\\
    \textit{Ireland}\\
    Email: 121442926@umail.ucc.ie $^1$, kguha@ucc.ie $^2$
  \end{minipage}
}
\maketitle

\begin{abstract}
Explainable {ML} algorithms are designed to provide transparency and insight into their decision-making process. Explaining how {ML} models come to their prediction is critical in fields such as healthcare and finance, as it provides insight into how models can help detect bias in predictions and help comply with GDPR compliance in these fields. {QML} leverages quantum phenomena such as entanglement and superposition, offering the potential for computational speedup and greater insights compared to classical {ML}. However, {QML} models also inherit the ‘black-box’ nature of their classical counterparts, requiring the development of explainability techniques to be applied to these {QML} models to help understand why and how a particular output was generated.

This paper will explore the idea of creating a modular, explainable {QML} framework that splits {QML} algorithms into their core components, such as feature maps, variational circuits (ansatz), optimizers, kernels, and quantum-classical loops. Each component will be analyzed using explainability techniques, such as {ALE} and {SHAP}, which have been adapted to analyse the different components of these {QML} algorithms. By combining insights from these parts, the paper aims to infer explainability to the overall {QML} model. 
\end{abstract}

\section{Introduction}
Machine learning has become an indispensable tool in recent years in many sectors, from healthcare diagnostics to financial market risk assessment. As this adoption grows, so does the requirement to understand ML predictions in human-interpretable terms, to help comply with GDPR regulations, and to interpret the black-box nature of these algorithms \cite{black-box}. Classical {ML} explainability techniques have been widely adopted to address this, but applying these techniques to their {QML} counterparts remain underdeveloped \cite{qml-explainability-limitations}. {QML} algorithms exploit quantum phenomena such as entanglement and superposition, and can help solve problems involving high-dimensional data, which would be infeasible or computationally intensive for classical {ML} \cite{qml-development}. This key core difference means that classical {ML} techniques may not be sufficient to explain how these models arrive at a prediction. This paper aims to adapt classical techniques to the core components of a {QML} algorithm, to allow for greater insight into the inner workings of these algorithms.
\subsection{Goal}
This paper introduces a modular framework for explainable {QML} algorithms that decomposes and divides them into interpretable components - feature maps, ansätze, optimizers and quantum kernels. The proposition is that explainability can emerge from analyzing each individual component of a {QML} algorithm, rather than analyzing trained models as a whole. The algorithm can be inferred as explainable by systematically applying explainability techniques to each of the core components. For comparison against the classical {ML}, classical data was used, encoded into quantum states using a feature map. By modularizing {QML} components, the paper aims are as follows:
\begin{itemize}
    \item \textbf{Component-Level Explainability:} Analyze how individual components contribute to the decision-making process of the model.
    \item \textbf{Modular Explainability Framework:} Create an analysis system to measure how different parts of the {QML} algorithms affect predictions.
    \item \textbf{Model Tuning Implications} Lay the groundwork for using techniques to help understand parts of the model that could be tweaked to improve performance.
\end{itemize}

\section{Background}
In {QML} algorithms, they take advantage of quantum phenomena to offer exponential computation power when compared to classical computing. In regards to superposition, where a qubit can exist in a state of zero, one, or a combination of both, this allows {QML} alforithms to perform multiple calculations on different data points or parameters at the same time, offering a speed increase \cite{qml-phenomena}. \\
By taking advantage of entanglement, where the state of one qubit can be affected by the state of another, this can be taken advantage of to represent correlations between different features in a dataset \cite{qml-phenomena}. This is because each feature is mapped to a single qubit, so for a dataset with $n$ features, it requires $n$ qubits.
\subsection{Quantum-Classical Algorithms}
This paper focused on the {VQC} and {QSVC} algorithms, which are both quantum-classical hybrid algorithms. This was primarily done due to \textit{quantum hardware limitations} as purely quantum {QML} models require large numbers of qubits and stable systems. Hybrid models only use quantum resources where they have the most impact. \cite{quantum-hardware}. Although there are some recent advances in regards to reducing error rates, such as Microsoft's Majorana 1 chip \cite{majorana-1}, there is still some time to go before these can be utilized to the level current hybrid models are used.
\subsection{Quantum Machine Learning Componentisation}
Decomposing {QML} algorithms offer several advantages for explainability insights when compared to monolithic approaches:
\begin{itemize}
    \item \textbf{Quantum-Classical Differences} \\
    Classical explainability techniques can make assumptions about a model that may not hold in quantum systems. By decomposing these algorithms, layer-wise techniques can be developed to help remove some of these assumptions that would other wise occur when analysed monolithicly. As quoted in \cite{qml-explainability-limitations2}, where it discusses: \begin{quote}
        'the fundamental differences between NNs and PQCs lead to incompatibilities between some classical explainability techniques and QML models, like the need to store intermediate states or clone information'
    \end{quote}
    \item \textbf{Layer-Wise Analysis} \\
    As mentioned above, a monolithic approach to explainability would make assumptions about a {QML} model that may not necessarily be correct. In saying this however gives us a problem: can a blanket approach be applied to all {QML} algorithms? \\
    The answer: no. Different algorithms have different structures, such as variational circuits and quantum kernels. This means that component based explainability techniques developed must be developed on specific components, rather than an algorithmic level.
    \item \textbf{Different Components, Different Importance} \\
    The component level analysis also provides insights into where different features may be important. For example, as realized in the evaluation chapter, different features may be important when initially encoding the classical data with the feature map, but those same features may not have much of an impact when training the quantum kernel. These kind of insights would be inherently lost when using a monolithic approach, potentially leading to cutting out features when trying to improve accuracy, as the monolithic approach may view a certain feature as not having that much importance towards the prediction, even though it had a large impact earlier in the algorithm.
\end{itemize}
\subsection{Feature Maps}
Feature maps are vital to {QML} algorithms. They are used in the first step when building {QML} models: transforming classical data into quantum states. These quantum states form the foundation for allowing {QML} algorithms to take advantage of quantum phenomena on the classical data. The feature map is \textit{model-agnostic}, meaning the encoded state produced by the feature map is the same for both models tested.
\\ There are three main types of feature map available via Qiskit:
\begin{itemize}
    \item \textbf{Basis Encoding} \\
    This encoding method performs a direct one-to-one mapping of classical bits to qubits. A bit with the value of $1$ would be mapped to a value of $|1\rangle$ and a bit of value $0$ would be mapped to a value of $|0\rangle$ \cite{encoding}. The advantages od this would be the simplicity and ease of implementation, with a major drawback being it can only encode simple data elements.
    \item \textbf{Angle Encoding} \\
    Includes feature maps such as Pauli feature map, Z and ZZ feature maps. These feature maps encode classical data values to rotation angles in quantum gates, leveraging the geometric representation of the bloch sphere, as seen in Figure \ref{fig:bloch-sphere}. \cite{encoding}
    \item \textbf{Amplitude Encoding} \\
    Includes circuits such as Raw Feature Vector. These circuits embed classical data into probability amplitudes of a quantum state. This offers exponential qubit efficiency and the cost of an increase in circuit depth. \cite{encoding}
\end{itemize}
\begin{figure}
\centering
\includegraphics[width=0.55\linewidth]{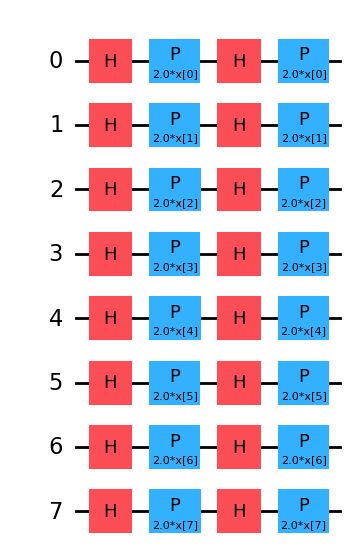}
\caption{\label{fig:z_feature_map}Circuit Diagram for a Z Feature Map with 2 repetitions.}
\end{figure}
\subsubsection{Z-Feature-Map}
The feature map chosen for this paper was the Z Feature Map, which is an angle-encoding feature map. The Z feature map is derived from the Pauli feature map, where the pauli strings are fixed as 'Z' \cite{z_feature_map}.  This first-order Pauli evolution circuit encodes input data $\vec{x} \in \mathbb{R}^n$, where n is the number of feature dimensions, as

\[
U_{\Phi(\vec{x})} = \exp \left( i \sum_{S \in I} \phi_{S}(\vec{x}) \prod_{i \in S} P_i \right)
\] 

Here, $S$ is a set of qubit indices that describes the connections in the feature map, $I$ is a set containing all these index sets, and $P_i \in \{ I, X, Y, Z \}$. Per default the data-mapping $\phi_S$ is

\[
\phi_{S}(\vec{x}) = 
\begin{cases} 
x_i \; \; \;  \text{if } S = \{i\} \\
\prod_{j \in S} (\pi - x_j) \; \; \; \text{if } |S| > 1
\end{cases}
\]
\cite{pauli_feature_map}.
\begin{figure}
\centering
\includegraphics[width=1\linewidth]{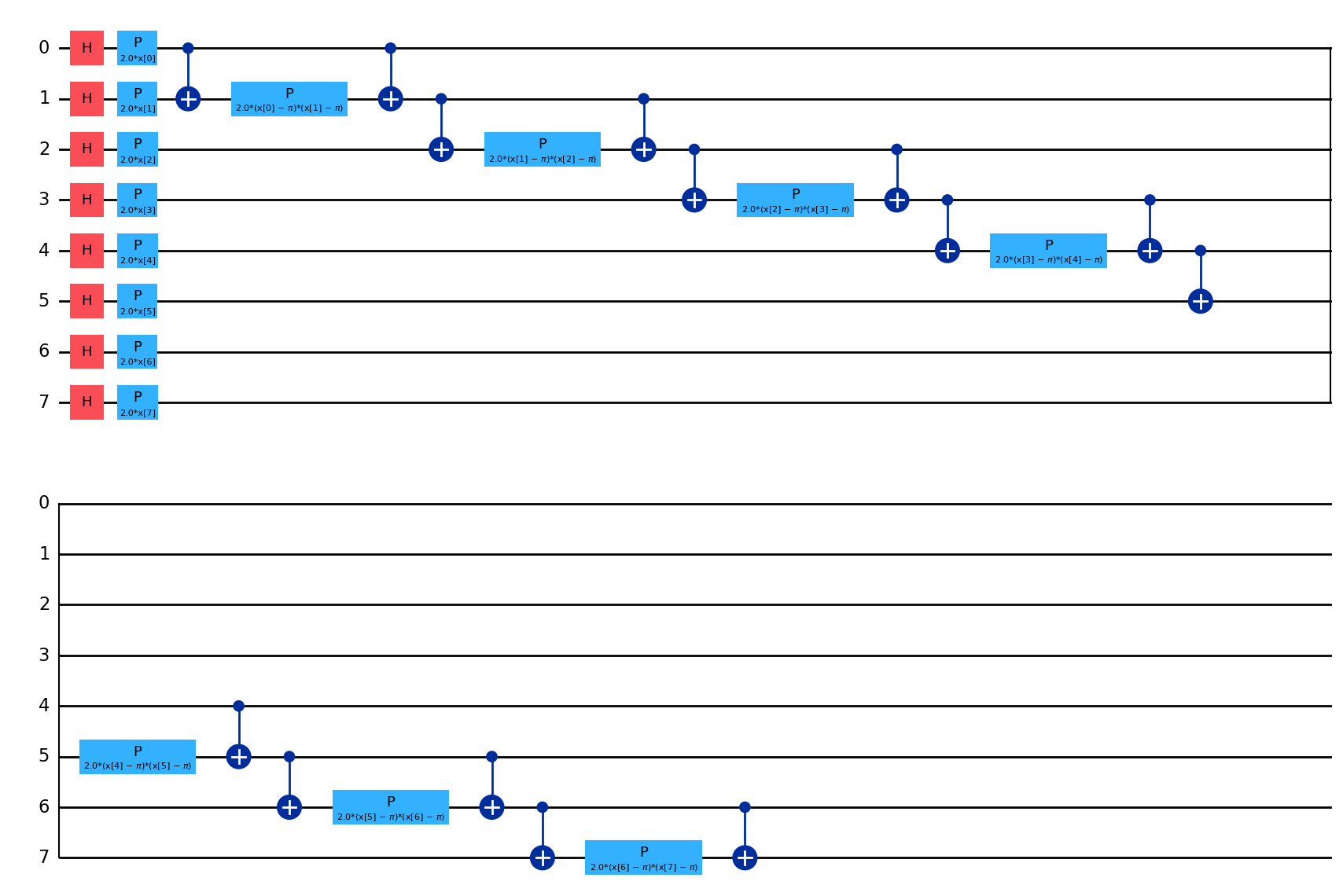}
\caption{\label{fig:zz_feature_map}Circuit Diagram for a ZZ Feature Map with linear entanglement and 1 repetition.}
\end{figure}
\\
This was implemented with two reps, as this was found to have the highest accuracy and F1 score when implementing the {VQC}. This feature map doesn't feature entanglement however, as shown in Figure~\ref{fig:z_feature_map}. Aside from being included in the best combination for the {VQC}, the other main advantage of using this was the computational time. As the Pima Indians Diabetes dataset has 8 features, this means that 8 qubits are needed when encoding. In testing, when an entangled feature map was used, this lead to significantly more computational overhead when training the models and computing explainability on the different components. Although this does not include entanglement, as shown in Figure~\ref{fig:z_feature_map} (there is no connection between the 8 qubits) the {VQC} and {QSVC} still offer a quantum advantage through their ansatz and quantum kernel respectively. \\
An example of a entanglement feature map can be seen in one of the ZZ feature maps tested; Figure~\ref{fig:z_feature_map}. This clearly shows the entanglement link between qubits through CNOT gates, where the '+' represents the target qubit and the 'dot' represents the control qubit.\\
In regards to implementation, the explainability framework was created with modularity in mind, so different feature maps can be dropped in and results can be still obtained.

\subsection{Variational Quantum Classifier (VQC)}
\begin{figure}
\centering
\includegraphics[width=1\linewidth]{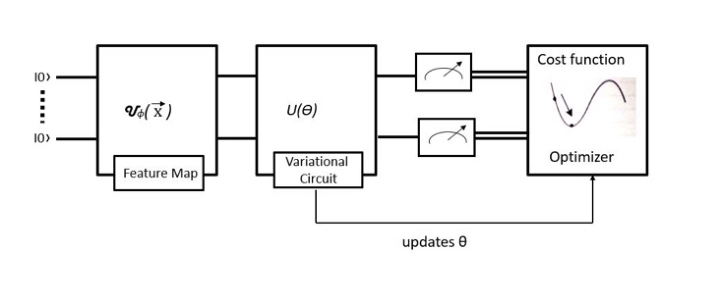}
\caption{\label{fig:vqc-workings}Diagram Showcasing how a {VQC} works. \cite{vqc-diagram-workings}}
\end{figure}

A {VQC} is a hybrid quantum-classical machine learning algorithm used for classification. It takes advantage of quantum phenomena, while utilizing classical optimization techniques \cite{vqc}. It works in a loop, as follows:
\begin{itemize}
    \item Classical data is encoded into quantum states using a feature map.
    \item The variational circuit (ansatz) manipulates these quantum states, and measurements are then taken from this circuit
    \item The measurements are fed into a classical optimizer, which adjusts the ansatz parameters to minimize some cost function
\end{itemize}
The ansatz parameters are updated iteratively until optimal parameters have been found. The {VQC} can then classify new data by encoding it, processing them through the circuit and analyzing the measurement results \cite{vqc}. The workings of a {VQC} are shown in Figure~\ref{fig:vqc-workings}.
\subsection{{VQC} Components}
\begin{figure}
\centering
\includegraphics[width=1\linewidth]{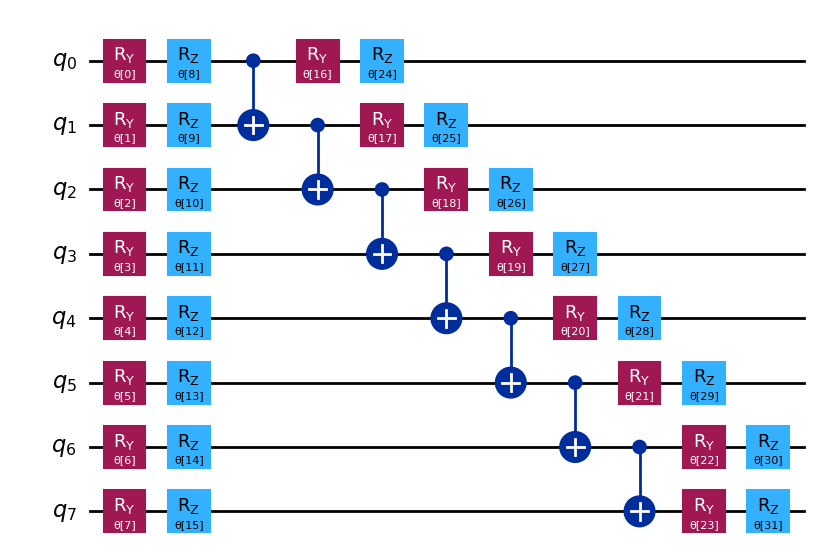}
\caption{\label{fig:eff_su2}Circuit Diagram for an Efficient SU2 ansatz with linear entanglement and 1 repetition.}
\end{figure}
\begin{figure}
\centering
\includegraphics[width=1\linewidth]{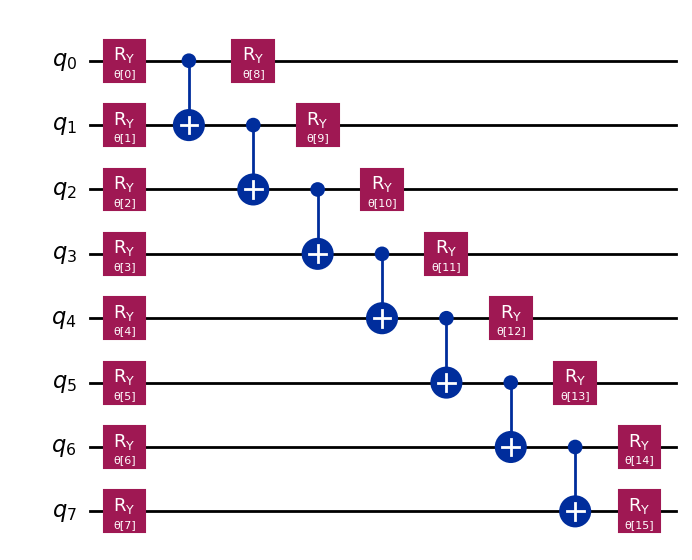}
\caption{\label{fig:real_amp}Circuit Diagram for a Real Amplitudes ansatz with linear entanglement and 1 repetition.}
\end{figure}
\subsubsection{Ansatz}
The ansatz is the {PQC} component of the {VQC}. It contains trainable parameters, represented by $\theta$. The ansatz applies a series of quantum gates with adjustable parameters to transform an input quantum state into an optimized quantum state \cite{variational-algorithms}.
There were 2 main types of ansatz circuits explored:
\begin{itemize}
    \item \textbf{Efficient SU2} \\
    The EfficientSU2 circuit consists of layers of single qubit operations spanned by SU(2) and $CX$ entanglements, as shown in Figure~\ref{fig:eff_su2}. This is a heuristic pattern that can be used to prepare trial wave functions for variational quantum algorithms or classification circuit for machine learning \cite{efficient-su2}.
    \item \textbf{Real Amplitudes} \\
    The Real Amplitudes circuit is a heuristic trial wave function that is used in chemistry applications or {QML} classification. It consists of alternating layers of $Y$ rotations and $CX$ entanglements, as shown in Figure~\ref{fig:real_amp}. It is called Real Amplitudes since the prepared quantum states will only have real amplitudes, the complex part is always 0 \cite{real-amplitudes}.
\end{itemize}
\subsubsection{Optimizer}
The optimizer component of the {VQC} is the classical proportion of the algorithm. When defining an optimizer in the {VQC}, an iteration is set. The optimizer aims to minimize a cost function to find the optimal parameters for the {PQC}. After the quantum state is passed through the ansatz, measurements are taken. The optimizer systematically scours the parameter space, and iteratively updates ansatz parameters to find the optimal values for the until it reaches convergence, or runs out of iterations. For binary classification, it aims to minimize cross-entropy loss by default. \\
The one chosen for the {VQC} implementation in this paper was \textbf{{COBYLA}}. {COBYLA} is a gradient-free optimizer, unlike others such as ADAM. This means it does not require gradient information, resulting in less overhead, as well as {COBYLA} requiring fewer circuits per evaluation when compared to gradient based optimizers. This was notable when testing other optimizers as gradient based optimizers were hit or miss if they were compatible with the {QML} algorithms.
\subsection{Quantum Support Vector Classifier (QSVC)}
A {QSVC} is a quantum-enhanced version of the traditional {SVC}. The {QSVC} uses a feature map to encode the classical data $\vec{x}_i$ into a high-dimensional quantum Hilbert space \cite{qsvc-101}, defined as $\psi(\vec{x})$. It takes advantage of quantum phenomena through its kernel, by computing the inner products of the quantum states encoded by the feature map: 
\[
K_{ij} \leftarrow | \langle \psi(\vec{x}_i), \psi(\vec{x}_j) \rangle |^2
\]

where K is the kernel matrix. This kernel matrix is then passed into a classical {SVC} optimizer, which calculates the optimal separation hyperplane in the quantum-enhanced feature space \cite{qml-implementation}.
\subsection{{QSVC} Components}
\subsubsection{Quantum Kernel}
The most intensive part of training a classical {SVC} is computing the kernel matrix. This is replaced with a quantum kernel in the {QSVC}. It essentially computes the similarity between two data points by calculating state fidelity between the quantum states. For each point $K_{ij}$ in kernel matrix $K$, the value corresponds to how similar the quantum states for data points $x_i$ and $x_j$ are. \\
As discussed before, the quantum kernel can capture complex data patterns by taking advantage of quantum phenomena that classical kernels cannot. \cite{q-kernel-tutorial}
\subsubsection{Classical {SVC}}
This part of the {QSVC} finds the decision boundaries based on the kernel matrix that has mapped the data into a higher dimensional space. This is called the 'kernel trick' \cite{q-kernel-tutorial}. It uses a subset of the training data - support vectors - to find the best decision boundary.
\subsection{Explainability Techniques}
As {QML} models gain more traction and development, the need for the creation of {XQAI} explainability techniques are also  utmost importance. There are many reasons as to why the development of these techniques are important:
\begin{itemize}
    \item \textbf{Model Trust and Transparency} \\
    When models have the ability to explain their decisions and predictions, users gain confidence in the accuracy and fairness of the models \cite{xai}.
    \item \textbf{Regulatory Compliance} \\
    Making models more transparent are essential to complying to regulatory requirements, such as GDPR requirements enforced in the European Union, which allows individuals to understand and challenge automated decisions. Explainability makes compliance easier by providing clear explanations as to how these decisions were made \cite{xai}.
    \item \textbf{Error Detection and Debugging} \\
    Explainable models allow engineers to trace back errors in predictions, allowing for more robust models. This error tracing is crucial in high-stakes environments, such as automated driving, where knowing why a system identified a pedestrian as a traffic sign would be crucial to trace and fix \cite{xai}.
    \item \textbf{Bias Mitigation} \\
    Models could potentially learn bias from training data, resulting in discriminatory decisions being made. Explainability techniques can address this by showing how a prediction was made, enabling the detection of these biases and laying the framework to remove them. 
\end{itemize}
For the development of these component-based explainability techniques for the {QML} models, \textit{model-agnostic} explainability techniques were the main focus for adaptation. The main reason for this was the flexibility of the implementation. Due to the adapted techniques being model-agnostic, this allowed for a consistent modified implementation of these techniques across all components.
\subsubsection{SHapley Additive eXplanations (SHAP)}
{SHAP} is an explainability method that is based on co-operative game theory. Provides a method for distributing a payout, which is typically the output of the model, among a coalition of players (input features) \cite{shap}. The classical goal of {SHAP} is to determine how much each feature contributes to the difference between the actual prediction and the average (central) prediction. \\
For a given prediction, {SHAP} calculates how much each feature moved the prediction away from the average prediction. The {SHAP} value for a given feature is then calculated by considering all possible combinations of features. Mathematically, the value is the weighted average of all possible differences when a feature is included versus excluded, which is represented by:
\[
\phi_i = \sum_{S \subseteq F \setminus \{i\}} \frac{|S|!\,(|F| - |S| - 1)!}{|F|!} \left[f_{S \cup \{i\}}(x_{S \cup \{i\}}) - f_S(x_S)\right]
\]
where:
\begin{itemize}
    \item \textbf{$\phi_i$} is the {SHAP} value for feature $i$
    \item \textbf{$F$} is the set of all features
    \item \textbf{$S$} is a subset of features
    \item \textbf{$f_S$} is a model trained on features in subset $S$
    \item \textbf{$x_S$} are the input values for features in $S$
\end{itemize}
\cite{shap}.
\subsubsection{Accumulated Local Effects (ALE)}
{ALE} is another explainability method, that focuses on feature values and how perturbed versions of feature values affect a models prediction by averaging the effects across multiple data instances to gain an understanding of feature impact. This was developed to over come the limitations of Partial Dependence Plots (PDPs), in which being they become unreliable when features are correlated \cite{ale_1}. \\
It works by dividing the range of each feature into bins calculates the difference in predictions when the feature value is perturbed from the lower bound to the upper bound of that bin. The differences in predictions are averaged across all instances that fall within each bin. This produces the 'local affect' for that specific bin of that feature. These local affects are then accumulated from lowest bin to the current one to create the {ALE} plot. Then these accumulated affects are centered by subtracting a constant, which ensures that the average effect throughout the feature range is zero \cite{ale-paper}. 

\[
= \int_{\textbf{z}_{0,S}}^{\textbf{x}_S} \left( \int_{\textbf{x}_C} \hat{f}^S (\textbf{z}_S, X_C) d\mathbb{P}(X_C|X_S = \textbf{z}_S) \right) d\textbf{z}_S - \text{constant}
\]

\cite{ale-paper}.
\subsection{Other Techniques}
Other techniques were also used in conjunction with the quantum adapted techniques above. These techniques did not involve the measurement of state fidelity, and are already widely used to help evaluate and understand different parts of classical portions of the algorithm, such as the evaluated kernel and the {SVC} decision function.
\subsubsection{Spectral Analysis}
Spectral analysis involves decomposing a kernel matrix into its eigenvalues and eigenvectors to uncover the underlying structure and intrinsic features of the data. By analysing the principal components using scree plots and PCA visualizations, it helpds clarify what underlying data patterns the kernel is using and gives a clearer understanding into the kernels behavior. 
\subsubsection{{SVC} Decision Function Visualisation}
This again uses {SHAP} to help gain a better understanding as to what the decision function is actually doing, and how features can impact it. It also helped in showing how different features in different parts of the algorithm impact differently.

\section{Previous QML Explainability Approaches}
There are several papers that explore the implementation of {QML} algorithms, and even multiple guides provided by IBM's Qiskit itself \cite{qml-tutorials}, but not many on making these algorithms explainable. This paper is the first to apply techniques to the core parts of {QML} algorithms, rather than viewing {QML} algorithms as monolithic entities and applying explainability techniques to the algorithm as a whole.
\subsection{'Study of Feature Importance for Quantum Machine Learning Models'}
This article was written by researchers at IBM, where they applied both Permutation Importance and {ALE} for feature importance on Fantasy Football data provided by ESPN \cite{fi-qml}. This again views the algorithms as monolithic entities, and does not show delve into component level specifics.
\subsection{'eXplainable AI for Quantum Machine Learning'}
A synthetic "bars and stripes" dataset was use in this study to highlight the effects of {SHAP} on three {PQC} classifiers - one qubit, two qubit and four qubit \cite{xai-qml}. This lacked comparisons with classical {ML} counterparts. To expand on this, a real dataset was used with more features, as well as testing against a wider range of {QML} algorithms.
\subsection{'Feature Importance and Explainability in Quantum Machine Learning'} This paper written in 2024 covered a wide and extensive range of explainability and feature importance techniques on both quantum and classical models, but used the Iris dataset, which does not translate well to real-world data, as well as viewing the algorithms as monolithic, rather than decomposing the algorithms into their core parts. \cite{luke-paper}
\subsection{'Explaining Quantum Circuits with Shapley Values: Towards Explainable Quantum Machine Learning'}
The article written by researchers in Germany explored the idea of assigning {SHAP} values to gates in a quantum circuit, but the framework used does not assess other critical components in {QML} algorithms, such as feature maps, kernel functions and optimization loops. The datasets used were also of a small scale, and the results found may not generalize well to real-world applications. The only method used for explainability in this study was also {SHAP} values, and does not explore other explainability techniques, which could reveal different insights into how a {QML} algorithm makes its prediction \cite{shap-paper}.
\subsection{'Explainable Heart Disease Prediction Using Ensemble-Quantum Machine Learning Approach'} This paper explores an ensemble-enhanced {QSVC}, trained on the Cleveland dataset. It also compares these results against other {QML} algorithms such as {VQC} as well as classical {ML} models. It also applies {SHAP} predictions to models, but again does not implement a comprehensive comparison and does not implement other explainability or feature importance methods to include in their comparisons \cite{heart-disease-paper}.
\\
\\
The main goal of this article, and how it differs from the ones listed above, is to provide a modular framework for explainability, while applying multiple explainability and feature importance methods to all of the core parts of a {QML} algorithm, on the Pima Indians Diabetes Dataset (binary classification). To evaluate all methods used a comprehensive and in depth analysis into the values will occur against classical {ML} algorithms.
lit review table

\section{Implementation}
As discussed earlier, classical explainability techniques typically focus on the model prediction. As the goal of this paper was to create component-level explainability techniques, these had to be adapted via the creation of 'pseudo models', which essentially act as a model but replaces the prediction with a state fidelity measure. Fidelity was chosen as a measure due to its ability to analyse quantum states. Classical, model-agnostic explainability techniques are oblivious to {QML} characteristics such as superposition and entanglement which inherently have an impact on a prediction. Using state fidelity in combination with classical techniques provide a more comprehensive overview as to what the model is actually doing. The implementation created takes copies of quantum states, put into the pseudo model to analyse what features are contributing to different class average quantum state.
\subsection{Quantum State Fidelity}
State fidelity was used as the measure to replace model performance when adapting explainability techniques. If $|\psi\rangle$ is the actual state and $|\phi\rangle$ is the target state, fidelity can be given as:
\[
F = |\langle\psi|\phi\rangle|^2
\] 
A fidelity of 1 indicates that the states are identical, and a fidelity of 0 indicates the states are orthogonal (different) \cite{state-fidelity}. This was accessible through a callable function available in Qiskit called \textit{state\_fidelity}, where it takes in two quantum states and outputs a fidelity value. \\
This was paramount to applying explainability techniques to different {QML} components, as these techniques are usually applied to the overall algorithm, and measure predictions. These were adapted to instead take in fidelity measurements.
\subsection{{SHAP} Adaptation}
In the adaptation of {SHAP} for the feature map and the {VQC} ansatz, helper functions for extracting the quantum state and measuring the similarity between two quantum states were created. These were then used in a lambda pseudo model to be passed into the {SHAP}'s libraries KernelExplainer, which is used to compute the SHAPley values, based on the pseudo model. The implementation takes in one sample, and then calculates how similar the quantum state is to the baseline diabetic and non-diabetic state. For the ansatz, SHAP values were computed on ansatz parameter importance. A new ansatz was created by slightly perturbing the original ansatz weights, and uses fidelity as a measurement of importance, and how these perturbed ansatz parameters affect the encoding process.\\
For applying {SHAP} to the quantum kernel, the kernel was iteratively evaluated compared to a central sample from each class. This was done to measure kernel fidelity - what features contribute to the similarity of the central sample of each class.
\subsection{{ALE} Adaptation}
The {ALE} adaptation was a little trickier. The original plan was to use the widely recognised \textit{alibi.explain} library for the {ALE} implementation, but this was unable to be used due to the lack of updates, and lead to conflicts as it was using earlier versions of numpy and SQLite. Although with more limited documentation, the package that had to be used for the {ALE} implementation was \textit{skexplain}. The process for implementation was similar to the {SHAP} implementation - using state fidelity - but the code implementation was drastically different. skexplain's ExplainToolKit expects an actual model with a defined 'predict' function. This lead to a different implementation of the pseudo model, as a predict function had to be defined. It still used the same helper functions to get the quantum state and state similarity however.

\section{Evaluation}
\subsection{Feature Map Explainability}
As the feature map is model-agnostic, this was the obvious choice for the first implementation of explainability techniques. The feature map used for both tested models was the Z Feature Map, whose circuit can be seen in Figure~\ref{fig:z_feature_map}. Two explainability techniques were adapted for the feature map - {SHAP} and {ALE}.
\subsubsection{{SHAP}}
This adapted the classical {SHAP} technique slightly by treating state fidelity prediction, but keeping features as the 'players'.
\begin{itemize}
    \item \textbf{Initial Encoded Quantum State Feature Sensitivity} \\
    The first explainability result was based on what features are important to the initial quantum state encoded by the feature map, regardless of class. It provides a local explanation for an input sample, and how much the values of each of the features in that sample contribute towards creating the encoded quantum state. This is shown in Figure~\ref{fig:general_fm_shap}, where Age plays the most significant role into defining the quantum state, as well as BloodPressure, SkinThickness and Pregnancies. This is denoted by the large, positive figures in the waterfall plot. This indicates that the feature map is expressive, as it is sensitive to input features, as is mapping inputs to different locations in the quantum feature space. This is a desirable property for classification, as it has the potential to be able to separate different data points. 
\begin{figure}
\centering
\includegraphics[width=1\linewidth]{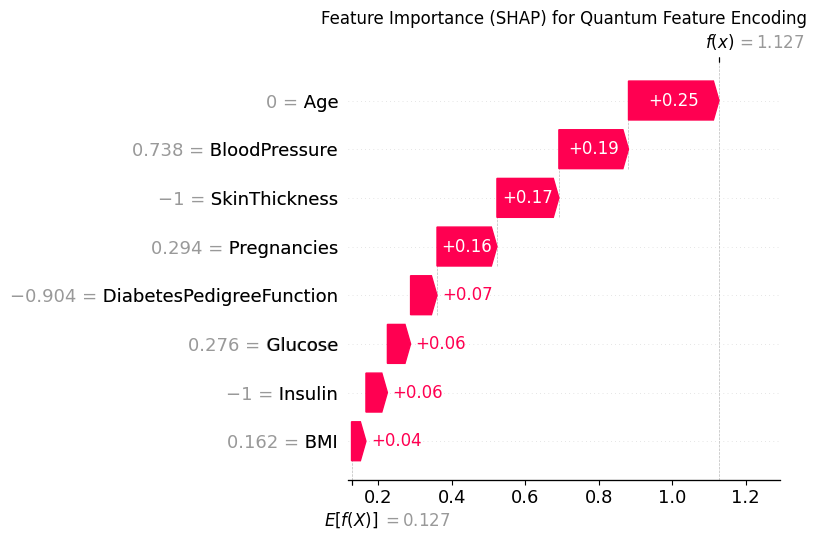}
\caption{\label{fig:general_fm_shap}{SHAP} Waterfall plot for features contributing to initial encoded quantum state.}
\end{figure}
    \item \textbf{Feature Importance for Class-Based Initial Encoded Quantum State} \\
    For further analysis, a class-based approach was implemented. This was a local explanation as to what features contribute to making the encoded quantum state look like each class. As seen in the Non-Diabetic waterfall plot in Figure~\ref{fig:non-diabetic_fm_shap}, we can see that Pregnancies, Glucose and Age all have positive values, meaning they are pushing away the samples encoded quantum state to that of the reference sample for the non-diabetic class. This is backed up in Figure~\ref{fig:diabetic_fm_shap}, where the same features are pushing it closer towards that of a diabetic sample. One thing to note is Diabetes Pedigree Function has positive values for both. This means that this is pushing away from typical examples of both classes in the quantum feature space. These techniques clearly show how localized explainability techniques can help with feature map tuning, and even help with feature engineering, when creating quantum models.
\begin{figure}
\centering
\includegraphics[width=0.9\linewidth]{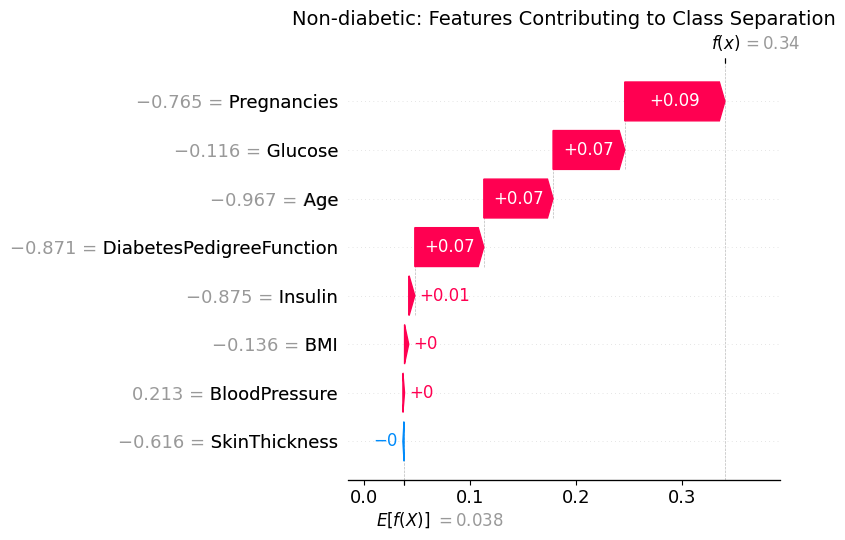}
\caption{\label{fig:non-diabetic_fm_shap}{SHAP} Waterfall showing what features push the quantum state towards the Non-Diabetic class.}
\end{figure}
\begin{figure}
\centering
\includegraphics[width=0.9\linewidth]{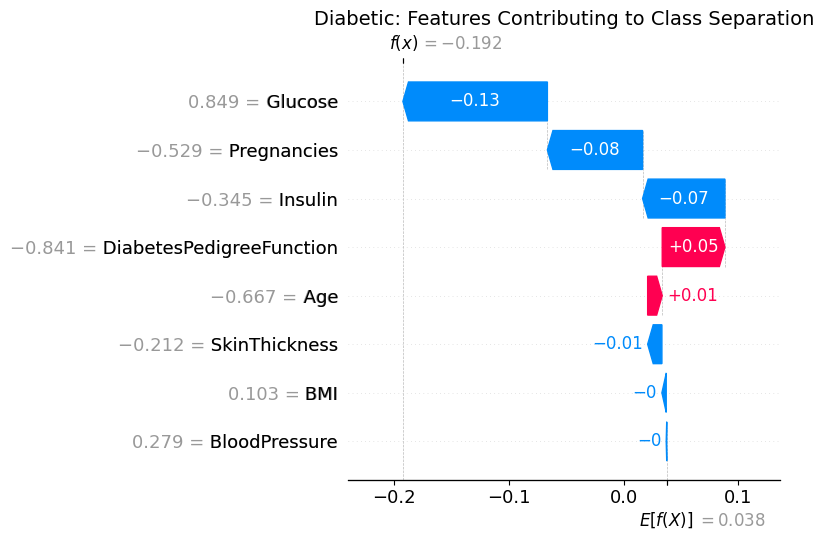}
\caption{\label{fig:diabetic_fm_shap}{SHAP} Waterfall showing what features push the quantum state towards the Diabetic class.}
\end{figure}
\end{itemize}

\subsubsection{{ALE}}
\begin{figure}
\centering
\includegraphics[width=0.9\linewidth]{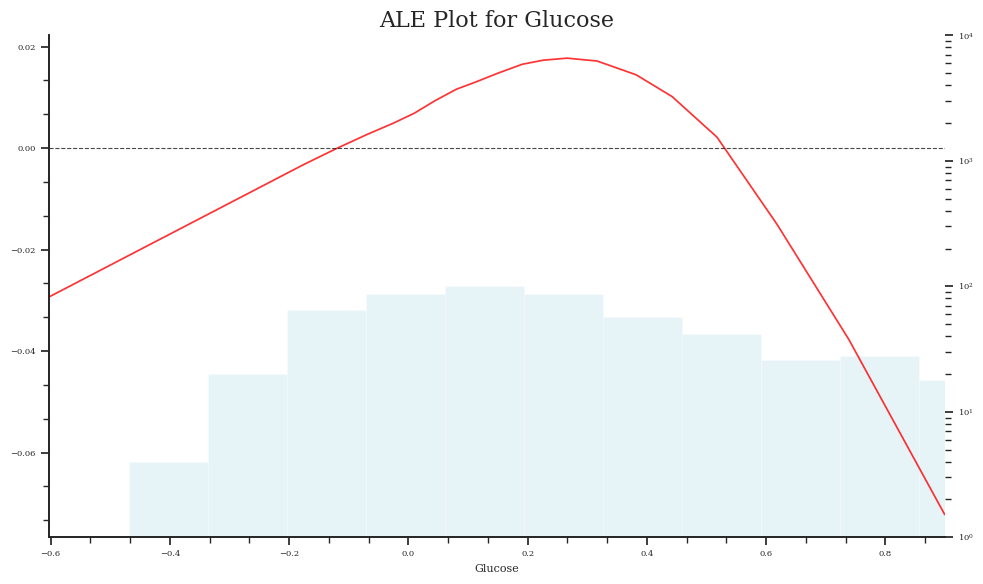}
\caption{\label{fig:ale_glucose}{ALE} plot for Glucose.}
\end{figure}

\begin{figure}
\centering
\includegraphics[width=0.9\linewidth]{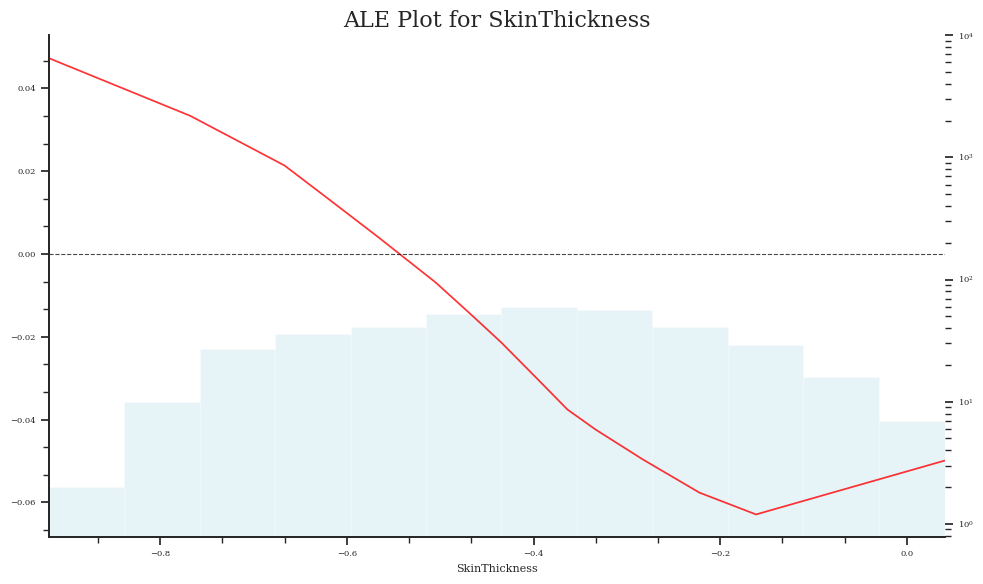}
\caption{\label{fig:ale_skin}{ALE} plot for SkinThickness.}
\end{figure}

The implementation for {ALE} was yet another local explanation based around one sample. It shows that when feature values are changed, how sensitive the feature map is to that change. This is relative to the average affect across the background data. For example, in Figure~\ref{fig:ale_glucose}, we can see it follows an 'n' shape. This means that due to the high {ALE} value when the scaled Glucose value is $\approx$ 0.2, the feature map is most sensitive to the value, for this sample. In ~\ref{fig:ale_skin}, we can see the feature map is most sensitive to low SkinThickness values, where as higher values make it less sensitive. This again can help show what feature ranges are most dominant, helping with model tuning and contributing to overall model explainability. 
\subsection{{VQC} Explainability}
This next section now focuses on {VQC} specific techniques.
\subsubsection{Ansatz}
\begin{figure}
\centering
\includegraphics[width=0.7\linewidth]{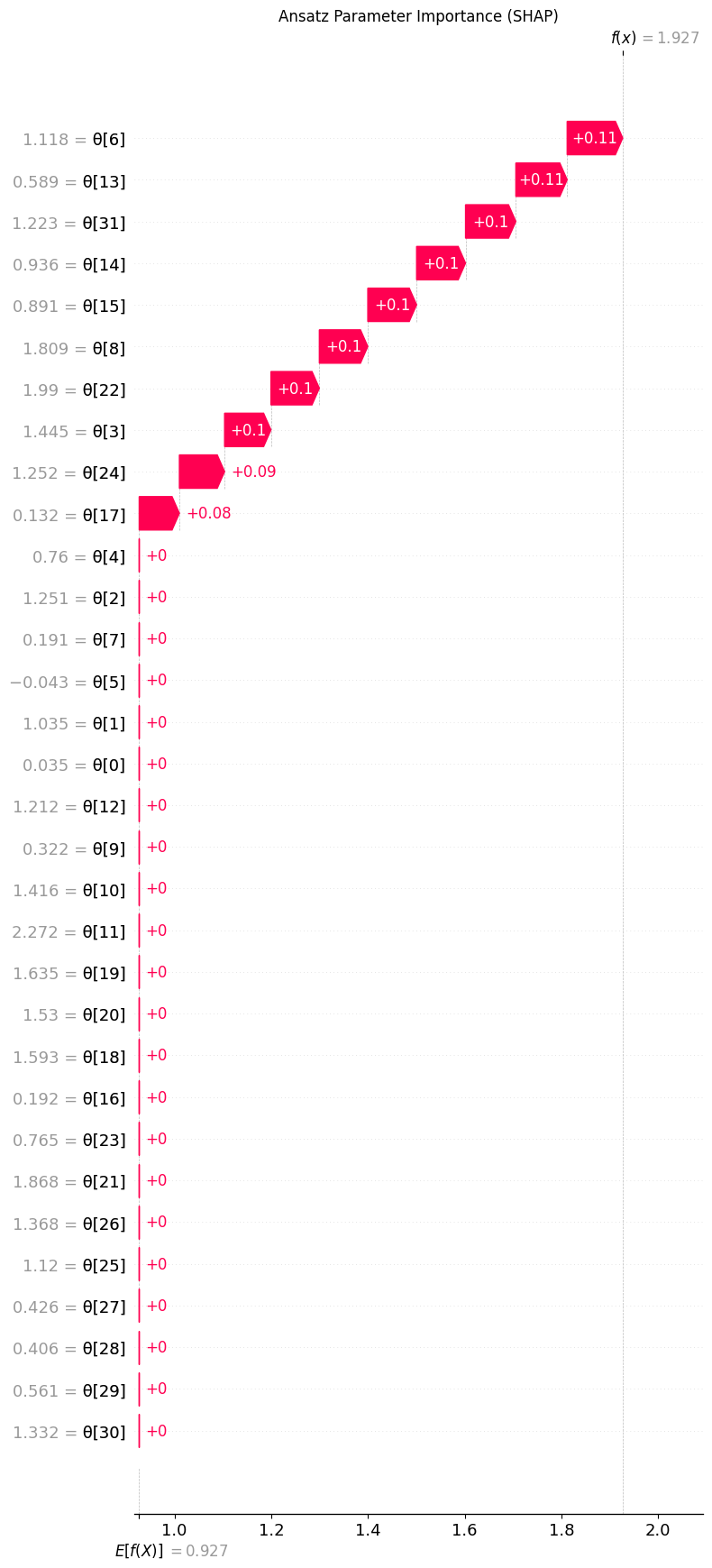}
\caption{\label{fig:ansatz_params}{SHAP} Waterfall plot showing parameter importance for the ansatz.}
\end{figure}

\begin{figure}
\centering
\includegraphics[width=1\linewidth]{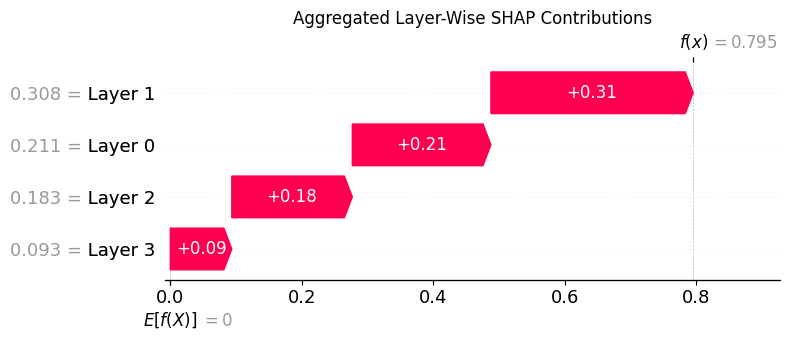}
\caption{\label{fig:ansatz_layers}{SHAP} Waterfall plot showing parameter layer importance for the ansatz.}
\end{figure}

The {SHAP} adaptation for the ansatz works by perturbing ansatz parameters, and measuring them against the trained ansatz from the {VQC}. It changes up the classical {SHAP} by changing the players to be the trainable ansatz parameters. By comparing the output state of the trained ansatz to the perturbed version, we can get a  {SHAP} value for each of the parameters, and show which ones are most important. This waterfall plot can be seen in Figure~\ref{fig:ansatz_params}, where we can see that 10 parameters have the most impact. This could be used with other techniques such as entanglement analysis to help isolate and fine tune the {VQC} on a circuit level. The ansatz used can be seen in Figure~\ref{fig:real_amp_full} (the one used had more repetitions). Analysing the parameter importance along side the entanglement that comes with this circuit can help remove the black box nature of the {VQC}. A layer-wise approach to analyse the parameters was also done in Figure~\ref{fig:ansatz_layers}, where it aggregated  the {SHAP} values for each layer (or repetition) to show which layer of parameters was most important.
\subsection{{QSVC} Explainability}
This next section now focuses on {QSVC} specific techniques.
\subsubsection{Quantum Kernel}
\begin{figure}
\centering
\includegraphics[width=1\linewidth]{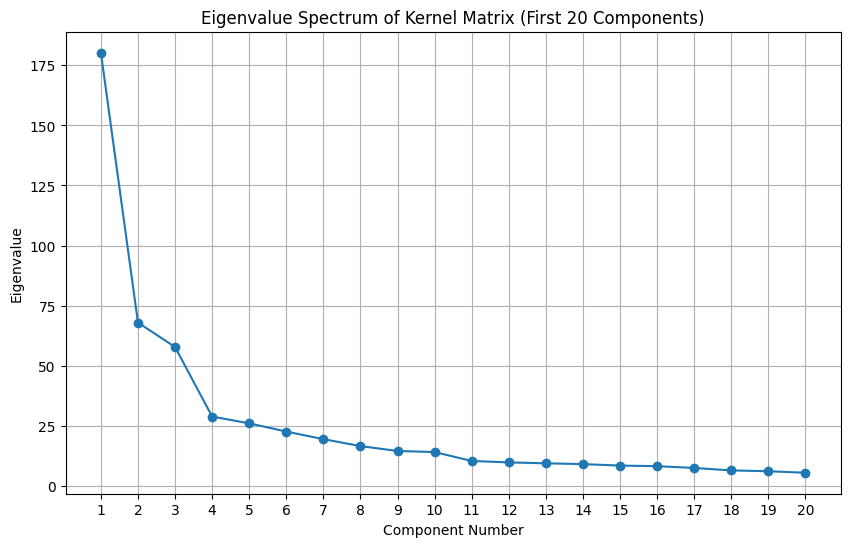}
\caption{\label{fig:spectrum}Eigenvalue Scree plot for first 20 components.}
\end{figure}

\begin{figure}
\centering
\includegraphics[width=1\linewidth]{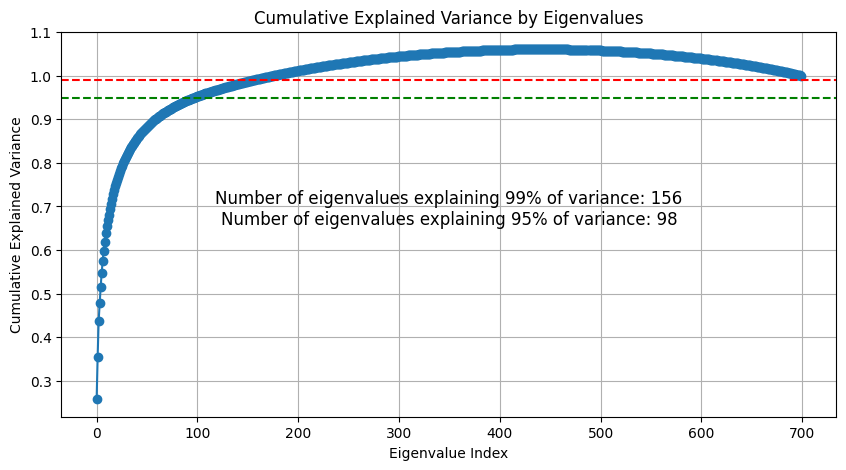}
\caption{\label{fig:cumsum_variance}Scatter plot for the cumulative explained variance.}
\end{figure}

\begin{figure}
\centering
\includegraphics[width=1\linewidth]{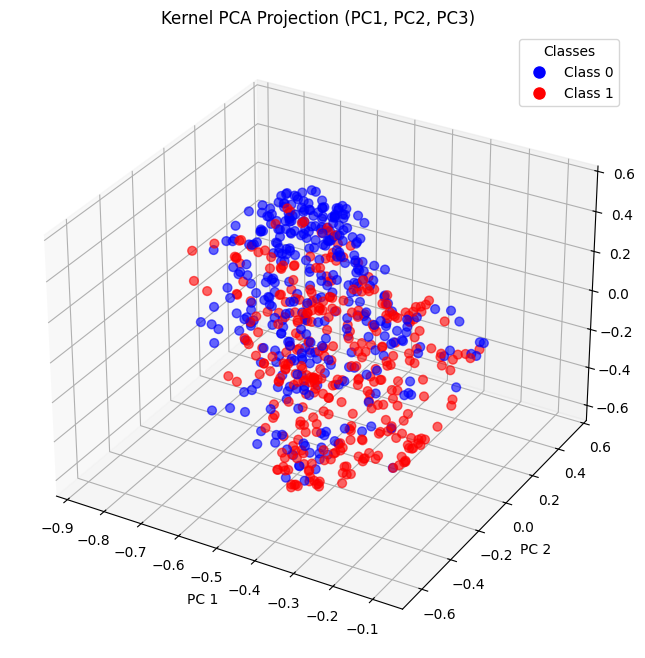}
\caption{\label{fig:pca_proj_123}PCA projection for principal components 1, 2, and 3.}
\end{figure}

\begin{figure}
\centering
\includegraphics[width=1\linewidth]{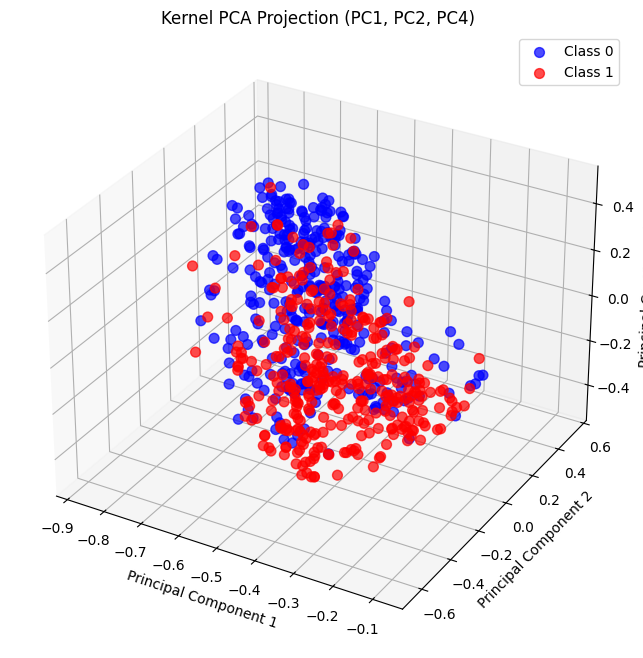}
\caption{\label{fig:pca_proj_124}PCA projection for principal components 1, 2, and 4.}
\end{figure}

\begin{figure}
\centering
\includegraphics[width=1\linewidth]{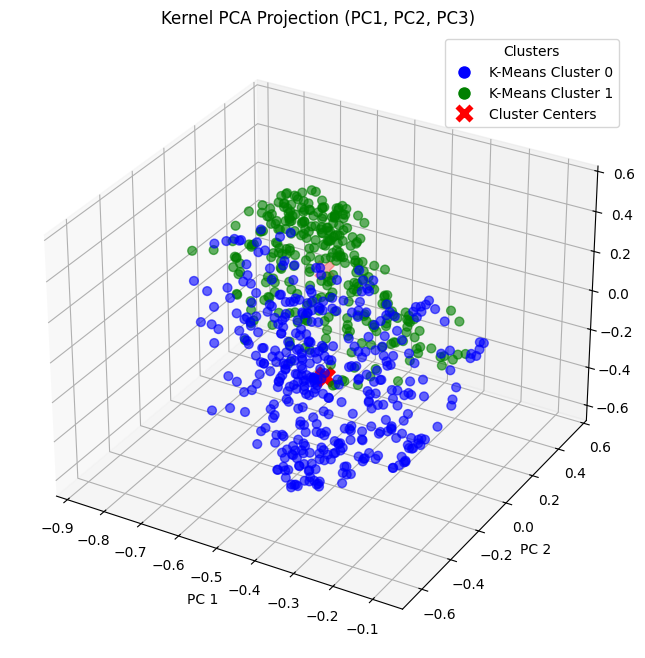}
\caption{\label{fig:pca_cluster}PCA projection for principal components 1, 2, and 3 showcasing clusters.}
\end{figure}

To help explain the Quantum Kernel, I based the analysis on eigenvalues derived from the computed kernel matrix. The eigenvalues were plotted in Figure~\ref{fig:cumsum_variance}, which shows that 156 components capture 99\% of the variance in the data, which implies a high effective dimensionality for the kernel. When this is compared to the total number of 700 components, this lays the groundwork for model simplification techniques, such as circuit simplification or attributing eigenvalues to model components. Figure~\ref{fig:spectrum} shows a plot of the first 20 eigenvalues, and highlights the dominant impact of the first four components. To analyse this further, PCA papers were created for principal component 1, 2, 3 (Figure~\ref{fig:pca_proj_123}) and 1, 2, 4 (Figure~\ref{fig:pca_proj_124}). Figure~\ref{fig:pca_proj_123} doesn't really show any distinct clusters for the classes, which is not great for classification. This is backed up when looking at the two distinct K-Means clusters in Figure~\ref{fig:pca_cluster}, where in cluster 0, 65\% of the data points belong to the diabetic class, and in cluster 1 69\% of the data points belong to the non-diabetic class, indicating the model is performing well but is still failing to encapsulate some of the underlying data patterns needed for classification. Figure~\ref{fig:pca_proj_124} shows a more separated cluster set, showing how the PCA paperion overtime captures more and more data variance. Using techniques like these can help again with model tuning, using eigenvalues to optimize circuits to help with further clarification. Model simplification techniques such as PCA reduction could also help prevent overfitting.
\subsubsection{Classical {SVC} Decision Function}
\begin{figure}
\centering
\includegraphics[width=0.9\linewidth]{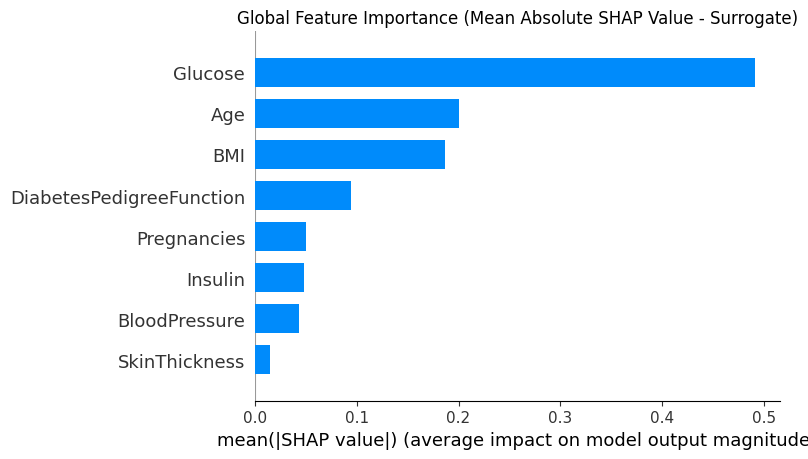}
\caption{\label{fig:shap_bar}{SHAP} bar plot showing global feature importance for the decision function.}
\end{figure}

\begin{figure}
\centering
\includegraphics[width=0.9\linewidth]{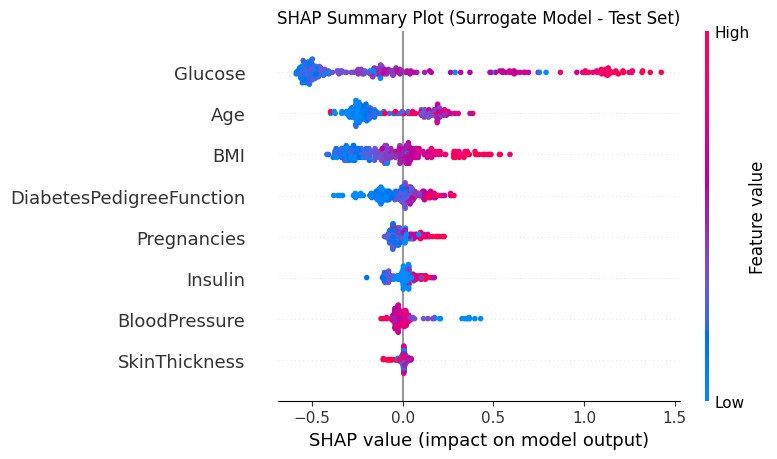}
\caption{\label{fig:shap_summary}{SHAP} summary plot for feature value impact on decision function.}
\end{figure}

To analyse the decision function, it was first computed by the {QSVC} on the test set. A surrogate model was then trained, using the input features but the {QSVC} decision function output as its target. This surrogate was trained until it got a near perfect replica of the outputs of the decision function (obtained a 0.98 $R^2$ score, so 98\% replication of results). This was able to capture {SHAP} values for each feature and their contribution to the decision function output, based on this surrogate. This can be seen in Figure~\ref{fig:shap_bar}, where it shows the global feature importance for each feature as approximated by the surrogate. We can see that Glucose had the most significant impact on the model, while also Age and BMI having strong impacts. SkinThickness can be seen not having that much of an impact on the decision function output. This can be compared to the results from Figure~\ref{fig:non-diabetic_fm_shap} and Figure~\ref{fig:diabetic_fm_shap}, where glucose has a significant impact in both {SHAP} plots for the feature map. SkinThickness, follows the same trend of not being that significant. Age and BMI being significant for decision function output and not for the initial encoded quantum state produced by the feature map is also something notable, as this counterintuitive interaction would be hidden had component level techniques not been implemented. This highlights that although Age and BMI had little impact in the initial encoding, they played a big role in the decision function. Likewise with pregnancies not being important in the decision function output, but had a significant impact in the encoding of the initial quantum state as shown in Figure~\ref{fig:general_fm_shap}. \\
The summary plot shown in Figure~\ref{fig:shap_summary} shows a more in depth global explanation as to the distribution of {SHAP} values and how they contribute towards the decision function output. This plot shows the distribution of feature values and how they contribute towards the output of the decision function. Higher {SHAP} values indicate the output is being pushed towards the diabetic class while lower {SHAP} values indicate the non-diabetic class. Blue dots indicate lower values for that feature while red dots indicate a higher feature value. This plot confirms real-world indicators, such as having higher glucose levels pushes the output more towards that of a diabetic patient. The plot also shows that higher BMI levels also push the output towards that of a diabetic. The horizontal spread indicates how much a feature can influence a decision, which can be contrasted with the earlier {SHAP} figures to show different feature importance at different parts of the algorithm. This plot could be extremely useful for visualizing how feature values impact decisions.

\section{Conclusion}
To answer the original question that this paper was based off:
\begin{center}
    'Can we make Quantum Machine Learning algorithms more explainable by applying explainability techniques to their core components rather than the overall algorithm?'
\end{center}

The answer: Yes. The component based explainability techniques developed in this project clearly show the benefits of using these techniques, as well as the need. QML algorithms, although built on top of classical Ml models (such as the {SVC}), the techniques adapted clearly show that in different parts of the algorithms, insights can be gained into feature and parameter importance that would otherwise be hidden by applying classical explainability techniques to the overall model. These techniques lay the groundwork for further exploration of quantum-native explainability techniques in the future, and it was significant accomplishment to develop, as of the writing of this report, the first component based explainability techniques to help further the emerging field of {XQAI}. It was extremely interesting to read the limited amount of papers in this field, and having the ability to be able to contribute to the ground works of {XQAI}. \\
It was extremely interesting to see, even with the limited size of the dataset and the training of modules on the simulator, some of the niche quirks with quantum machine learning, such roughly equal performance on a smaller subset of data and the ability to not generalize well to imbalanced datasets.
\bibliographystyle{unsrturl}  
\bibliography{references}

\begin{thebibliography}{10}

\bibitem{black-box}
Please, explain. interpretability of black-box machine learning models, 2019.
\newblock URL: \url{https://www.appsilon.com/post/please-explain-black-box}.

\bibitem{qml-explainability-limitations}
E.~Gil-Fuster, J.~R. Naujoks, G.~Montavon, T.~Wiegand, W.~Samek, and J.~Eisert.
\newblock Opportunities and limitations of explaining quantum machine learning.
\newblock {\em arXiv preprint}, 2024.
\newblock URL: \url{https://arxiv.org/html/2412.14753v1}.

\bibitem{qml-development}
K.~Najafi, S.~F. Yelin, and X.~Gao.
\newblock The development of quantum machine learning.
\newblock {\em Harvard Data Science Review}, 2022.
\newblock \href {https://doi.org/10.1162/99608f92.5a9fd72c}
  {\path{doi:10.1162/99608f92.5a9fd72c}}.

\bibitem{qml-phenomena}
Sumit Asthana.
\newblock Use of superposition and entanglement in quantum algorithms.
\newblock URL:
  \url{https://www.linkedin.com/pulse/use-superposition-entanglement-quantum-algorithms-sumit-asthana}.

\bibitem{quantum-hardware}
KqantumG Research Labs~Pvt Ltd.
\newblock Hybrid quantum-classical models: Unlocking the power of quantum
  machine learning, 2024.
\newblock URL:
  \url{https://www.linkedin.com/pulse/hybrid-quantum-classical-models-unlocking-j0mxf}.

\bibitem{majorana-1}
Chetan Nayak.
\newblock Microsoft unveils majorana 1, the world’s first quantum processor
  powered by topological qubits, 2025.
\newblock URL:
  \url{https://azure.microsoft.com/en-us/blog/quantum/2025/02/19/microsoft-unveils-majorana-1-the-worlds-first-quantum-processor-powered-by-topological-qubits/?msockid=0e634514af3c6aef0cd050d3aed66b64}.

\bibitem{qml-explainability-limitations2}
E.~Gil-Fuster, J.~R. Naujoks, Wojciech Samek, and Jens Eisert.
\newblock Opportunities and limitations of explaining quantum machine learning.
\newblock {\em arXiv preprint}, 2024.
\newblock URL: \url{https://arxiv.org/html/2412.14753v1}.

\bibitem{encoding}
Siddhartha Sharma and Renugadevi N.
\newblock Survey of encoding techniques for quantum machine learning.
\newblock {\em Cybernetics and Physics Vol. 13}, 2024.
\newblock URL:
  \url{https://inspirehep.net/files/7a1f4eae2f5fadb4c1ea45dd0b884584}.

\bibitem{z_feature_map}
IBM.
\newblock Z feature map documentation.
\newblock URL:
  \url{https://docs.quantum.ibm.com/api/qiskit/qiskit.circuit.library.z_feature_map}.

\bibitem{pauli_feature_map}
IBM.
\newblock Pauli feature map documentation.
\newblock URL:
  \url{https://docs.quantum.ibm.com/api/qiskit/qiskit.circuit.library.PauliFeatureMap}.

\bibitem{vqc}
H.~Qi et~al.
\newblock Variational quantum algorithms: fundamental concepts, applications
  and challenges.
\newblock {\em Quantum Information Processing}, 2024.
\newblock \href {https://doi.org/10.1007/s11128-024-04438-2}
  {\path{doi:10.1007/s11128-024-04438-2}}.

\bibitem{vqc-diagram-workings}
Aaron Baughman, Daniel Bohm, Micah Forster, Eduardo Morales, Jeff Powell, Shaun
  McPartlin, Raja Hebbar, and Kavitha Yogaraj.
\newblock Large scale diverse combinatorial optimization: Espn fantasy football
  player trades.
\newblock 11 2021.
\newblock \href {https://doi.org/10.48550/arXiv.2111.02859}
  {\path{doi:10.48550/arXiv.2111.02859}}.

\bibitem{variational-algorithms}
IBM~Quantum Learning.
\newblock Variational algorithm design.
\newblock URL:
  \url{https://learning.quantum.ibm.com/course/variational-algorithm-design/variational-algorithms}.

\bibitem{efficient-su2}
IBM.
\newblock Efficientsu2, 2025.
\newblock URL:
  \url{https://docs.quantum.ibm.com/api/qiskit/qiskit.circuit.library.EfficientSU2}.

\bibitem{real-amplitudes}
IBM.
\newblock Realamplitudes, 2025.
\newblock URL:
  \url{https://docs.quantum.ibm.com/api/qiskit/qiskit.circuit.library.RealAmplitudes}.

\bibitem{qsvc-101}
S.~Faldu.
\newblock Quantum support vector machine 101, 2023.
\newblock URL:
  \url{https://dzone.com/articles/quantum-support-vector-machine-101}.

\bibitem{qml-implementation}
S.~Navneet and S.~Raj~Pokhrel.
\newblock An independent implementation of quantum machine learning algorithms
  in qiskit for genomic data.
\newblock {\em arXiv preprint}, 2024.
\newblock URL: \url{https://arxiv.org/pdf/2405.09781/}.

\bibitem{q-kernel-tutorial}
IBM.
\newblock Quantum kernel machine learning, 2025.
\newblock URL:
  \url{https://qiskit-community.github.io/qiskit-machine-learning/tutorials/03_quantum_kernel.html}.

\bibitem{xai}
BotPenguin.
\newblock Explainable ai: Benefits and challenges, 2024.
\newblock URL: \url{https://botpenguin.com/glossary/explainable-ai}.

\bibitem{shap}
Scott~M. Lundberg and Su-In Lee.
\newblock A unified approach to interpreting model predictions.
\newblock {\em Conference on Neural Information Processing Systems}, 2017.

\bibitem{ale_1}
Oracle.
\newblock Accumulated local effects, 2020.
\newblock URL:
  \url{https://accelerated-data-science.readthedocs.io/en/v2.6.3/user_guide/model_explainability/accumulated_local_effects.html}.

\bibitem{ale-paper}
Daniel~W. Apley and Jingyu Zhu.
\newblock Visualizing the effects of predictor variables in black box
  supervised learning models.
\newblock {\em Journal of the Royal Statistical Society Series B: Statistical
  Methodology}, 82, 2020.
\newblock \href {https://doi.org/10.1111/rssb.12377}
  {\path{doi:10.1111/rssb.12377}}.

\bibitem{qml-tutorials}
IBM.
\newblock Machine learning tutorials - qiskit machine learning 0.8.2.
\newblock URL:
  \url{https://qiskit-community.github.io/qiskit-machine-learning/tutorials/index.html}.

\bibitem{fi-qml}
A.~Baughman, K.~Yogaraj, R.~Hebbar, S.~Ghosh, R.~U. Haq, and Y.~Chhabra.
\newblock Study of feature importance for quantum machine learning models.
\newblock {\em arXiv preprint}, 2022.
\newblock URL: \url{https://arxiv.org/pdf/2202.11204}.

\bibitem{xai-qml}
P.~Steinmuller, T.~Schulz, F.~Graf, and D.~Herr.
\newblock explainable ai for quantum machine learning.
\newblock {\em arXiv preprint}, 2022.
\newblock URL: \url{https://arxiv.org/pdf/2211.01441}.

\bibitem{luke-paper}
L.~Power and K.~Guha.
\newblock Feature importance and explainability in quantum machine learning.
\newblock {\em arXiv preprint}, 2024.
\newblock URL: \url{https://arxiv.org/abs/2405.08917v1}.

\bibitem{shap-paper}
R.~Heese.
\newblock Explaining quantum circuits with shapley values: Towards explainable
  quantum machine learning.
\newblock {\em Quantum Machine Intelligence}, 2025.
\newblock \href {https://doi.org/10.1007/s42484-025-00254-8}
  {\path{doi:10.1007/s42484-025-00254-8}}.

\bibitem{heart-disease-paper}
G.~Abdulsalam, S.~Meshoul, and H.~Shaiba.
\newblock Explainable heart disease prediction using ensemble-quantum machine
  learning approach.
\newblock {\em Intelligent Automation and Soft Computing}, 2022.
\newblock \href {https://doi.org/10.32604/iasc.2023.032262}
  {\path{doi:10.32604/iasc.2023.032262}}.

\bibitem{state-fidelity}
QuEra.
\newblock Qubit fidelity, 2025.
\newblock URL: \url{https://www.quera.com/glossary/fidelity}.

\end{thebibliography}

\end{document}